\documentstyle[aps,prd]{revtex}
\begin{document}
\title{Thin-shell wormholes: Linearization stability}
\author{Eric Poisson\cite{Eric-e-mail} and Matt Visser\cite{e-mail}}
\address{Physics Department, Washington University, St. Louis,
         Missouri 63130-4899, USA}
\date{June 1995}
\twocolumn[
\maketitle
\parshape=1 0.75in 5.5in
The class of spherically-symmetric thin-shell wormholes provides
a particularly elegant collection of exemplars for the study of
traversable Lorentzian wormholes. In the present paper we consider
linearized (spherically symmetric) perturbations around some assumed
static solution of the Einstein field equations. This permits us to
relate stability issues to the (linearized) equation of state of
the exotic matter which is located at the wormhole throat.
\vskip 0.125 in
\parshape=1 0.75in 5.5in
PACS number(s): 04.20.-q    04.20.Gz    04.25.-g    04.90.+e
]

\bibliographystyle{prsty}
\narrowtext
\section{INTRODUCTION}

The class of thin-shell wormholes~\cite{Visser89a,Visser89b} provides
a particularly elegant collection of exemplars for the study of
traversable Lorentzian wormholes~\cite{Morris-Thorne,MTY}. Within
the class of spherically-symmetric thin-shell wormholes, the dynamics
of the wormhole is completely specified (up to overall integration
constants such as the total mass of the wormhole system) once an
equation of state is specified for the ``exotic matter'' which is
located at the wormhole throat~\cite{Visser89b}.

The stability of such wormholes has previously been considered for
certain specially chosen equations of state~\cite{Visser89b,Visser}.
This type of analysis addresses the question of stability in the
sense of proving bounded motion for the wormhole throat; no static
solution to the wormhole equations need exist.

In the present paper we analyse the stability of spherically-symmetric
thin-shell wormholes by considering linearized radial perturbations
around some assumed static solution of the Einstein field equations.
This permits us to rephrase stability issues in terms of constraints
on the equation of state of the ``exotic matter''. Our analysis
follows closely that of Ref.~\cite{Brady-Louko-Poisson}.

The two types of stability analyses are complementary and permit
the extraction of somewhat different information regarding the
wormhole systems.

\section{Schwarzschild surgery}

To construct the wormholes of interest, apply the by now standard
cut-and-paste construction:  Take {\em two} copies Schwarzschild
spacetime, and remove from each manifold the four-dimensional
regions described by
\begin{equation}
\Omega_{1,2} \equiv \{ r_{_{1,2}} \leq a \;|\; a > 2M \},
\end{equation}
where $a$ is a constant.
The resulting manifolds have boundaries
given by the timelike hypersurfaces
\begin{equation}
\partial\Omega_{1,2} \equiv \{ r_{_{1,2}} = a \;|\; a> 2M \}.
\end{equation}
Now {\em identify} these two timelike hypersurfaces ({\sl
i.e.},\ $\partial\Omega_1 \equiv \partial\Omega_2$).  The resulting
manifold {$\cal M$} is geodesically complete and possesses two
asymptotically flat regions connected by a traversable Lorentzian
wormhole.  The throat of the wormhole is at $\partial\Omega$.

Because {$\cal M$} is piecewise Schwarzschild, the Einstein
tensor is zero everywhere {\it except} at the throat, where it
is formally singular: the Einstein tensor is a Dirac distribution
on the manifold. Using the field equations, the surface
stress-energy tensor can be calculated in terms of the jump
in the second fundamental form across $\partial \Omega$. This
prescription, known as the ``thin-shell formalism''~\cite{Israel66},
is standard. We now review it briefly.

Adopt Gaussian normal coordinates at the throat.  Let $\eta$ denote
proper distance away from the throat (in the normal direction),
with $\eta$ positive in $\Omega_1$ and negative in $\Omega_2$.
The second fundamental forms are then
\begin{equation}
K^i{}_j{}^\pm =
{1\over2} \; g^{ik}
\left.{\partial g_{kj}\over\partial\eta}\right|_{\eta=\pm0};
\end{equation}
they are functions over the surface $\partial \Omega$.
The Ricci tensor at the throat is easily calculated in terms of the
discontinuity in the second fundamental forms.  Define $[K_{ij}] =
K_{ij}{}^+ - K_{ij}{}^-$. Then~\cite{Visser89b,Visser}
\begin{equation}
R^\mu{}_\nu  =
 \left( \matrix{ [K^i{}_j] & 0\cr0&[K] \cr} \right)
 \delta(\eta),
\end{equation}
where $[K]$ denotes the trace of $[K^i{}_j]$.
This, together with the Einstein field equations,
implies that the stress-energy tensor is
localized at the throat:
\begin{equation}
T^{\mu\nu} = S^{\mu\nu} \; \delta(\eta),
\end{equation}
with
\begin{equation}
S^i{}_j =
-{c^4\over8\pi G}
\left( [K^i{}_j] - \delta^i{}_j \; [K] \right).
\end{equation}
Writing the surface stress-energy tensor in terms of the surface
energy density $\sigma$, and surface pressure $p$, one has
\begin{equation}
S^i{}_j = \mbox{diag}(-\sigma,p,p).
\end{equation}
Now specializing to spherical symmetry and adopting units
such that $G=c=1$, the thin-shell equations become
\begin{equation}
\sigma =
-{1\over4\pi} [K^\theta{}_\theta], \qquad
p =
+{1\over8\pi}
\left( [K^\tau{}_\tau] + [K^\theta{}_\theta] \right).
\end{equation}

To analyse the dynamics of the wormhole, we permit the radius of the
throat to become a function of time, $a \mapsto a(t)$. A simple
computation then yields~\cite{Visser89b,Visser}
\begin{eqnarray}
[K^\theta{}_\theta] &=& \frac{2}{a}
\sqrt{1-2M/a+\dot{a}^2}, \\
{[K^\tau{}_\tau]} &=& \frac{\ddot{a}+M/a^2}{\sqrt{1-
2M/a+\dot{a}^2}}.
\end{eqnarray}
Note that $\dot a$ denotes $da/d\tau$, where the parameter $\tau$
measures proper time along the wormhole throat.

The Einstein field equations reduce to
\begin{eqnarray}
\sigma &=& -{1\over2\pi a} \sqrt{1-2M/a+{\dot a}^2}; \\
p      &=& +{1\over4\pi a}
          { 1 -M/a + {\dot a}^2 + a \ddot a \over
             \sqrt{1-2M/a+{\dot a}^2}}.
\end{eqnarray}
It is easy to check that these imply energy conservation:
\begin{equation}
\frac{d}{d\tau}\, \sigma {\cal A} +
p \frac{d}{d\tau}\, {\cal A} = 0,
\end{equation}
where ${\cal A} = 4\pi a^2$.
In this equation, the first term corresponds to a change in
the throat's internal energy, while the second term corresponds
to the work done by the throat's internal forces.

\section{Stability Analysis}

The Einstein equations obtained in the previous section
may be recast as
\begin{equation}
{\dot a}^2 - 2M/a - (2\pi\sigma a)^2 = -1;
\end{equation}
\begin{equation}
\dot \sigma = -2 (\sigma+p) {\dot a \over a}.
\end{equation}
If we choose a particular equation of state, in the form $p
= p(\sigma)$, then we can formally integrate the conservation
equation and obtain
\begin{equation}
\ln(a) = -{1\over2} \int {d\sigma\over\sigma+p(\sigma)}.
\end{equation}
This relationship may then be formally inverted to yield $\sigma$
as a function of the wormhole radius: $\sigma = \sigma(a)$. Once
this is done, the first Einstein equation can be written in the
form
\begin{equation}
{\dot a}^2 = -V(a);
\qquad
V(a) = 1 -2M/a -[2\pi\sigma(a) a]^2.
\end{equation}
This single dynamical equation completely determines the motion
of the wormhole throat.

One may now choose to investigate particular equations
of state, as is done for instance in~\cite{Visser89b,Visser}.
Alternatively, one might attempt a more general analysis. We have
found that a particularly simple, though still instructive, choice
is to consider linearized fluctuations around an assumed static
solution characterized by the constants $a_0$, $\sigma_0$, and
$p_0$. Note that these constants are (by assumption) inter-related:
\begin{eqnarray}
\sigma_0 &=& -{1\over2\pi a_0} \sqrt{1-2M/a_0}; \\
p_0      &=& +{1\over4\pi a_0}
                 {1-M/a_0 \over \sqrt{1-2M/a_0}}.
\end{eqnarray}

We now insert these relations into the dynamical equation,
expanding to second order in $a-a_0$. Generically we
would have
\begin{eqnarray}
V(a)
&=&  V(a_0) + V'(a_0) [a-a_0]
\nonumber\\
&&\mbox{} + {\textstyle \frac{1}{2}}
V''(a_0) [a-a_0]^2 + O([a-a_0]^3),
\end{eqnarray}
where a prime denotes $d/da$.
However, because we are linearizing around a static solution at
$a=a_0$, we know that $V(a_0)=0$, and $V'(a_0)=0$. To leading
order, therefore, $V(a) = \frac{1}{2} V''(a_0) [a-a_0]^2$.

To compute the various derivatives, it is useful to rewrite
the conservation equation as
\begin{equation}
[\sigma(a) a ]' = -(\sigma+2p).
\end{equation}
Differentiating once more,
\begin{eqnarray}
[\sigma(a) a ]''
&=& -(\sigma' + 2p')
\nonumber\\
&=& -\sigma'\left(1+2{\partial p\over\partial\sigma}\right)
\nonumber\\
&=& 2 \left(1+2{\partial p\over\partial\sigma}\right)
{\sigma+p\over a}.
\end{eqnarray}
We now {\em define} a parameter $\beta$ by the relation
\begin{equation}
\beta^2(\sigma) \equiv
\left.{\partial p\over\partial\sigma}\right|_{\sigma}.
\end{equation}
The physical interpretation of $\beta$ is a matter of some subtlety
which we shall subsequently discuss. For now, we simply consider
$\beta$ to be a useful parameter related to the equation of state.
Using this definition,
\begin{equation}
[\sigma(a) a ]'' =
2 \left(1+2\beta^2\right)
{\sigma+p\over a}.
\end{equation}
Collecting the preceding results, we obtain
\begin{equation}
V'(a) = {2M\over a^2} +
8\pi^2 \sigma a \; \left(\sigma + 2 p \right),
\end{equation}
and
\begin{eqnarray}
V''(a) &=& -{4M\over a^3}
- 8\pi^2 \Bigl[ ( \sigma + 2 p )^2
\nonumber\\ & & \mbox{}
+2 \sigma \left(1+2\beta^2\right)
\left(\sigma+p\right)
\Bigr].
\end{eqnarray}

When evaluated at the static solution $a=a_0$ these equations
yield the expected results $V(a_0)=0$, and $V'(a_0) = 0$.
Furthermore,
\begin{eqnarray}
V''(a_0) &=& -2 a_0^{-2}
\Bigg[ {2M\over a_0} + {M^2/a_0^2 \over 1-2M/a_0}
\nonumber\\
&&\mbox{}
+ \left(1+2\beta_0^2\right)\left(1-{3M \over a_0}\right)
\Bigg].
\label{E-Vpp}
\end{eqnarray}
The equation of motion for the wormhole throat is, at this order
of approximation,
\begin{equation}
{\dot a}^2 = - {\textstyle \frac{1}{2}}
V''(a_0) [a-a_0]^2 + O([a-a_0]^3).
\end{equation}
So for $V''(a_0)>0$ the wormhole is stable, while for $V''(a_0)<0$
perturbations can grow (at least until the nonlinear regime
is reached). Thus the wormhole is stable if and only if:
\begin{equation}
{2M\over a_0}
+ {M^2/a_0^2 \over 1-2M/a_0}
+ (1+2\beta_0^2)\left(1-{3M\over a_0}\right) < 0.
\end{equation}
We shall now study this equation in detail.

If one treats $a_0$ and $M$ as specified quantities, stability may be
rephrased as a restriction on the parameter $\beta_0$:
\begin{equation}
\beta_0^2\left(1-{3M\over a_0}\right) <
-{1-3M/a_0 + 3(M/a_0)^2\over 2(1-2M/a_0)}.
\end{equation}
The RHS of this inequality is always negative, while the LHS flips
sign at $a_0=3M$. One deduces
\begin{equation}
\beta_0^2 < -{1-3M/a_0 + 3(M/a_0)^2\over 2(1-2M/a_0)(1-3M/a_0)};
\qquad a_0>3M.
\end{equation}
\begin{equation}
\beta_0^2 > -{1-3M/a_0 + 3(M/a_0)^2\over 2(1-2M/a_0)(1-3M/a_0)};
\qquad a_0<3M.
\end{equation}
The region of stability, in the $(\beta_0^2)$--$(a_0/M)$ plane,
is depicted in Fig.~\ref{fig1}. Note that as we have formulated
the problem, it is meaningless to ask what
happens for $a_0 < 2M$.

On the other hand, if one treats $\beta_0$ as an externally specified
quantity, stability  may be rephrased as a restriction on the the
allowable radius ($a_0$) of the assumed static solution. The boundary
of the region of stability is given by the curve $V''(a_0/M;\beta_0)=0$.
This gives a quadratic equation for $a_0/M$,
\begin{equation}
3(1+4\beta_0^2) \left({M \over a_0}\right)^2  -
(3+10\beta_0^2) \left({M \over a_0}\right)  +
    1+2\beta_0^2 = 0,
\end{equation}
with roots
\begin{equation}
a_0^\pm =
{ 6 (1+4\beta_0^2) M \over
3 + 10\beta_0^2 \mp \sqrt{4\beta_0^4-12\beta_0^2-3} }.
\end{equation}
The discriminant is real only for $\beta_0^2 \geq {3\over2} +
\sqrt{3} \approx 3.23205$, and for $\beta_0^2 \leq {3\over2} -
\sqrt{3} \approx -0.23205$.  After a bit of fiddling to make sure
that one is in the physically relevant region, the regions of
stability are given by
\begin{eqnarray}
&I:& \qquad \beta_0^2 \geq {3\over2} + \sqrt{3},
\qquad
a_0^- < a_0 < a_0^+ ;
\\
&II:& \qquad \beta_0^2 \leq -1/2,
\qquad \quad
a_0 > a_0^-.
\end{eqnarray}
For $ -\frac{1}{2} < \beta_0^2 < \frac{3}{2}+\sqrt{3}$,
the wormhole is unstable for all values of $a_0$. Of course,
this is just a reformulation of the result previously
discussed, as plotted in Fig.~\ref{fig1}.

\section{Conclusion}

Under normal circumstances one might (naively) try to interpret
$\beta_0$ as the speed of sound in the exotic matter located at
the wormhole throat. Furthermore, under normal circumstances one
would require $\beta_0$ to lie in the interval $\beta_0\in (0,1]$.
(One would normally deduce $\beta_0>0$ from the assumed stability
of matter, and argue that $\beta_0\leq1$ based on the requirement
that the speed of sound should not exceed the speed of light.) If we
restrict the speed of sound to lie in this standard range, then a
glance at the figure shows that spherically-symmetric thin-shell
wormholes are always unstable to linearized radial perturbations.
Naively therefore, insisting on stability seems possible only at
the cost of requiring somewhat perverse restrictions on the speed
of sound.

A certain level of caution in this interpretation is in order: To
start with, we are already dealing with ``exotic matter,''
in the sense that the surface energy density $\sigma_0$ is
negative. Thus, naive arguments depending on the stability of
matter and the positivity of energy should be taken with a grain
of salt.  In particular, the usual proof that $\beta_0 > 0$ is here
completely side stepped. Our analysis indeed shows that once the
effect of the wormhole's gravitational field is included, stability
implies that large traversable wormhole ($a_0 > 3M$) are stable
only for $\beta_0^2 <0$!

[There are several known examples of such exotic $\beta^2 <0$ behavior:
In the test field limit, the Casimir vacuum between parallel plates
is known to be of the form $T^{\mu \nu} \propto
\mbox{diag}(-1,1,1,-3)$. By integrating over the region
between the plates, the three-dimensional surface
stress-energy takes the form $T^{ij} \propto
\mbox{diag}(-1,1,1)$. In this case
$\beta^2 = \partial\sigma/\partial p = -1$. A similar
argument shows that $\beta^2=-1$ for false vacuum, for which
$T^{\mu\nu} = \Lambda g^{\mu \nu}$, where $\Lambda$ is a constant.]

Furthermore, the interpretation of $\beta_0$ as the speed of sound
is itself problematic insofar as one does not have a detailed
microphysical model for the exotic matter. This is simply not
available within the confines of the present analysis. Although
$\beta_0$ has the dimensions of a speed, and although one might
expect $\beta_0$ to be of the same order of magnitude as
the speed of sound, there is no guarantee that $\beta_0$
actually {\it is} the speed of sound. A detailed microphysical
model for the exotic matter would be necessary in order to settle
such issues with any certainty. We therefore conclude that wormhole
configurations with $|\beta_0^2| > 1$ should not be ruled out
{\it a priori}.

We have studied the linearization stability of a thin, spherically
symmetric, traversable wormhole against radial perturbations.
We have explicitly presented the region of
stability in terms of the mass of the wormhole $M$, the radius of
the wormhole throat $a_0$, and a parameter $\beta_0$ related to
the equation of state. Although the region of stability lies in
a somewhat unexpected region of the $(\beta_0^2)$--$(a_0/M)$ plane,
stable wormholes might nevertheless be physically acceptable. To
decide for sure would require a detailed microphysical model for
the exotic matter; this lies outside the scope of the present paper.

\acknowledgements

This research was supported by the U.S. National Science Foundation
under Grant No.~PHY 92-22902 and the U.S. Department of Energy.


\begin{figure}
\caption[]{Regions of stability: Traversable wormholes in the indicated
regions of the $\beta^2_0$ versus $a_0/M$ plane are stable against radial
perturbations. The higher dashed line is the curve $\beta_0^2 =
\frac{3}{2} + \sqrt{3}$; the lower dashed line is the curve
$\beta_0^2 = -\frac{1}{2}$.}
\label{fig1}
\end{figure}


\begin{references}
\bibitem[*]{Eric-e-mail}Electronic mail: poisson@wuphys.wustl.edu
\bibitem[\dag]{e-mail}Electronic mail: visser@kiwi.wustl.edu
\bibitem{Visser89a} M. Visser, Phys. Rev. D {\bf 39}, 3182 (1989).
\bibitem{Visser89b} M. Visser, Nucl. Phys. {\bf B328}, 203 (1989).
\bibitem{Morris-Thorne} M.S. Morris and K.S. Thorne, Am. J. Phys.
    {\bf 56}, 395 (1988).
\bibitem{MTY} M.S. Morris, K.S. Thorne, and U. Yurtsever, Phys.
    Rev. Lett. {\bf 61}, 1446 (1988).
\bibitem{Visser} M. Visser, {\it Lorentzian wormholes --- from
    Einstein to Hawking\/} (American Institute of Physics, New
    York, 1995).
\bibitem{Brady-Louko-Poisson} P.R. Brady, J. Louko, and E.
    Poisson, Phys. Rev. D {\bf 44}, 1891 (1991).
\bibitem{Israel66} N. Sen, Ann. Phys. (Leipzig) {\bf 73},
    365 (1924); K. Lanczos, Ann. Phys. (Leipzig) {\bf 74},
    518 (1924); W. Israel, Nuovo Cimento {\bf 44B}, 1
    (1966); erratum---{\it ibid.} {\bf 48B}, 463 (1967).
\end{references}
\end{document}